*Research Article*

# Regression Analysis of Ordinal Panel Count Data in Recurrent Medication Non-adherence

Jiangjie Zhou and Baosheng Liang *

Department of Biostatistics, School of Public Health, Peking University.

*Correspondence should be addressed to the following author:
Baosheng Liang (Dr.)
Department of Biostatistics, School of Public Health, Peking University

Address: Room 401, Public Health Building, #38 Xueyuan Rd., Haidian District, Beijing 100191, P.R China
Email: liangbs@hsc.pku.edu.cn

## Abstract

Panel count data arise in clinical trials when patients are asked to report their occurrences of events of interest periodically but the exact event times are unknown, only the count of events between two successive examinations are observed. Ordinal panel count data goes even further as the exact event counts are not observed, the only information available is rank of event counts, for example, 'never', 'sometimes' and 'always'. Currently, there is lacking of standard and efficient methods for analyzing this type of data. In this paper, we proposed a semiparametric proportional intensity model to analyze such data. We developed a maximum sieve likelihood estimation using monotone spline under the nonhomogeneous Poisson process model assumption for statistical inference. Simulation studies show that our method performs well with finite sample sizes and is relatively robust to model misspecification. In addition, we compared the proposed method with other competitors and the proposed method outperforms in various settings. Finally, we investigated the recurrence of medication non-adherence in a clinical trial on non-psychotic major depressive disorder using the proposed method.

**Keywords:** recurrent events; ordinal panel count data; proportional intensity model; medication non-adherence

## 1 Introduction

In many clinical trials, patients may be asked to report the occurrences of events of interest between two successive examination. However, it is hard for patients to recall the exact times and numbers of events. As a result, only rank of event count between two ex-aminations, for example, `never`, `sometimes` and `always`, is observed. One motivating example of this work comes from studying medication adherence in a Sequenced Treatment Alternatives to Relieve Depression (STAR*D) study[1]. Researchers are interested in evaluating the effects of covariates, such as age, gender and race on medication adherence, which is important for the effectiveness of treatments. In STAR*D study, patients received anti-depression drugs and were asked to report times of missing medicine, including 9 levels, such as `never missing medicine` and ` rarely missing medicine `, since their last visits. The fact that the exact times and counts of events were not observed poses significant challenges to regression analysis, both computational and theoretical. Another example is a survey on drug abuse among American high school students, where respondents cannot recall the exact number of drug abuse but can only report the level of usage [2]. Other examples include the severity of toxicity from tumor chemotherapy [3] and the intensity of immune reactions after organ transplants [4].

We term these recurrent events data with partially observed information as ordinal panel count data. They share similarity with traditional panel count data where the times to events of interest is never observed[5] and ordinal data which is produced by discretization of a continuous latent variable [6]. Many models have been suggested to analyze both types of data. For panel count data, researches have been focused on dealing with multivariate panel count data[7–9], informative censoring/examination[10–12], relaxing Poisson process assumptions using estimating equations[10,13] and variable selection[14,15]. For ordinal data analysis, many classical models exist, such as proportional odds model based on cumulative probability of latent variable and continuation ratio model based on conditional probability[6]. A generalization of above model to longitudinal data was made possible by introducing cumulative link mixed effects model[16,17].

In recent years, some researchers try to incorporate ordinal regression models into the field of recurrent event analysis[3], via a joint-modeling approach, but they only consider scenarios where exact event times are available. Liang et al. took unobserved event times into ac-count, but their method only applies to a binary response case[18]. Currently, there is lacking of efficient and valid method for analyzing ordinal panel count data. Usually, people treat the ordinal response as count response and use standard panel count analysis procedure such as the augmented estimating equations (AEE)[10]. Another way is to treat the response as longitudinal ordinal response and then use cumulative link mixed effects model (CLMM)[19] for longitudinal ordinal data. However, potential bias could be introduced for these methods, which could lead to incorrect inference.

In this paper, we propose an efficient method to analyze ordinal panel count data. We assume that the event process follows a proportional intensity non-homogenous Poisson process model, and use a cumulative probability approach to model the ordinal response. For estimating model parameters, we employ a spline-based sieve maximum likelihood estimation procedure and an optimization trick to solve the vanishing gradients problem caused by the discrete nature of Poisson distribution.

The remainder of this paper is organized as follow. In Section 2, we proposed the model and inference method with baseline cumulative intensity function using a sequence of I-splines in the cases with known or unknown cut points. Section 3 reports numerical results from extensive simulation studies. Section 4 illustrates the proposed method with an application of analyzing non-adherence medication data in the STAR*D study. A discussion concludes in Section 5.

## 2  Model and Inference Method

Let $N_i(t), i = 1, \ldots, n$ denote the total number of recurrent events before time $t$ for the $i$th patient, and $X_i$ be the time-independent covariate that relates to $N_i(t)$. To study the effect of $X$ on the recurrent event of interest with given covariate $X_i$, we assume that $N_i(t)$ follows a non-homogeneous Poisson process with intensity function

$$\lambda_i(t|X_i) = \lambda_0(t) \exp(\beta^T X_i), \qquad (1)$$

where $\beta \in \mathbb{R}^p$ is a p-dimensional regression parameter of interest, and $\lambda_0(t)$ is an unspecified baseline hazard function. $\Lambda_0(t) = \int_0^t \lambda_0(u) du$ is the cumulative intensity function or mean function.

Let $0 = T_{i0} < T_{i1} < \cdots < T_{im_i} \leq \tau$ be the visit time points for subject $i$, where $m_i$ is the total number of observation (or visit) time points for subject $i$ and $T_{im_i}$ is the last time point (censoring time) when subject $i$ is lost to follow-up, $\tau$ is the maximum observation time for this study. Let $\Delta_{ij} = N_i(T_{i,j}) - N_i(T_{i,j-1})$ be count of events that have occurred in the interval $(T_{i,j-1}, T_{ij}]$. We assume $(T_{i1}, \ldots, T_{im_i})$ are independent of $N_i(T_{i1}), \ldots, N_i(T_{im_i})$ given the covariates. Thus, the censoring mechanism is conditionally independent of the event processes. $n_{ij}$ is what we observe in panel count data. However, in our setting, we only observe the ranks or orders of $n_{ij}$, we denote this variable by $Y_{ij}$, where

$$Y_{ij} = \begin{cases} 1, & \Delta_{ij} \leq \gamma_1 \\ 2, & \gamma_1 < \Delta_{ij} \leq \gamma_2 \\ \vdots \\ K, & \Delta_{ij} > \gamma_{K-1} \end{cases}, \qquad (2)$$

and $\boldsymbol{\gamma} = (\gamma_1, \ldots, \gamma_{K-1})$ is cut points in the ordinal definition of $Y_{ij}$. For simplicity, we define $\gamma_0 = -1$ and $\gamma_K = \infty$.

## 2.1 The case with known cut points

In a simple setting, $\boldsymbol{\gamma}$ is known, for example, in a substance abuse study[2], cut points for occurrences of violence behavior are specified by the researchers. In this case, we can easily derive the likelihood function under a Poisson assumption and model (1)

$$L_n(\boldsymbol{\beta}, \Lambda_0) = \prod_{i=1}^n \prod_{j=1}^{m_i} \prod_{k=1}^K p_{ijk}^{I(Y_{ij}=k)}, \tag{3}$$

where $p_{ijk}$ is the probability that $n_{ij}$ equals $k$,

$$p_{ijk} = Pr(Y_{ij} = k | X_i, \boldsymbol{\gamma}) = Pr(\gamma_{k-1} < \Delta_{ij} \leq \gamma_k)$$
$$= \exp\{-\mu_{ij}(\beta, \Lambda)\} \left( \sum_{\Delta_{ij}=\gamma_{k-1}+1}^{\gamma_k} \frac{[\mu_{ij}(\beta, \Lambda)]^{\Delta_{ij}}}{\Delta_{ij}!} \right),$$

where $\mu_{ij}(\beta, \Lambda_0) = \int_{T_{i,j-1}}^{T_{i,j}} \exp(\beta^T X_i) \, d\Lambda_0(t)$. Let $F_\lambda(x)$ be the cumulative distribution function of Poisson distribution with mean parameter $\lambda$. Then an equivalent expression for $p_{ijk}$ is

$$p_{ijk} = F_{\mu_{ij}(\beta,\Lambda)}(\gamma_k) - F_{\mu_{ij}(\beta,\Lambda)}(\gamma_{k-1}).$$

Our goal is to estimate $(\beta, \Lambda_0)$ by maximizing $L_n(\beta, \Lambda_0)$ over the parameter space $\Theta = \mathscr{B} \times \mathscr{A}$, where $\mathscr{B}$ is a bounded open subset of $\mathbb{R}^p$, and

$$\mathscr{A} = \{ \Lambda_0 : \Lambda_0 \text{ is a nondecreasing function over } [0, \tau] \}.$$

We develop a monotone spline-based sieve maximum likelihood estimation. Specifically, we use I-spline[20] functions to model the baseline mean function $\Lambda_0(t)$: let the set of spline knots be $0 = t_1 = \cdots = t_l < t_{l+1} < \cdots < t_{m_n+l} < t_{m_n+l+1} = \cdots = t_{m_n+2l} = \tau$, where $m_n$ is knots number depending on n, and $l$ is the order of I-spline. Let $\{I_i(\cdot), i = 1, \cdots, m_n + l\}$ be the I-spline basis functions corresponding to these knots. Then, we can approximate $\Lambda_0(t)$ by a linear combination of these basis functions:

$$\Lambda_0(t) \approx \sum_{i=1}^{m_n+l} \alpha_i I_i(t),$$

where $\alpha_i \geq 0, i = 1, \cdots, m_n + l$. To drop the constraints on $\alpha_i$, we can write $\alpha_i$ as $\tilde{\alpha}_i^2$ and use these unconstrained $\tilde{\alpha}_i$ as parameters. This optimization problem can be solved using the Broyden-Fletcher-Goldfarb-Shanno (BFGS) algorithm[21]. We denote the maximum likelihood estimator by $(\hat{\beta}, \hat{\alpha})$.

For the convenience of statistical inference, we propose to estimate the covariance matrix by a sandwich estimator

$$\hat{V} = \left\{ D_{h_n}^2 pl(\hat{\beta}) \right\}^{-1} \sum_{i=1}^{n} D_{h_n} pl(\hat{\beta}) D_{h_n} pl(\hat{\beta})^T \left\{ D_{h_n}^2 pl(\hat{\beta}) \right\}^{-1}, \quad (4)$$

where $pl(\hat{\beta})$ is the profile log-likelihood function, $D_{h_n}$ and $D_{h_n}^2$ are the first-order and second-order numerical derivatives with perturbation constant $h_n \sim cn^{-1/2}$. Specifically, for $\xi \in \mathbb{R}^p$,

$$D_{h_n} f(\xi) = \left\{ \frac{f(\xi + h_n e_i) - f(\xi)}{h_n} \right\}_{i=1,\dots,p},$$

And

$$D_{h_n}^2 f(\xi) = \left\{ \frac{f(\xi) - f(\xi + h_n e_i) - f(\xi + h_n e_j) + f(\xi + h_n e_i + h_n e_j)}{h_n} \right\}_{i,j=1,\dots,p},$$

where $e_i$ is the $i$th canonical vector of $\mathbb{R}^p$. We then estimate variance of $\hat{\Lambda}_0(t)$ by

$$Var\left(\hat{\Lambda}_0(t)\right) = \sum_{l=1}^{L} I_l(t)^2 Var(\alpha_l). \quad (5)$$

## 2.2 The case with unknown cut points

In the case of cut points $\gamma = (\gamma_1, \dots, \gamma_{K-1})$ being unknown, it is almost impossible to optimize (3) using regular algorithms based on derivatives, because equation (3) is a discrete function of the parameter $\gamma$, so gradients vanish for $\gamma$. It is a nonlinear mixed integer programming problem and generally hard to solve.

To deal with this optimization problem, we introduce an optimization trick. First, define a continuous version of Poisson distribution, which has a distribution function of $\tilde{F}_\lambda(x)$,

$$\tilde{F}_\lambda(x) = \begin{cases} 0, & x \leq 0, \\ \frac{\Gamma(x, \lambda)}{\Gamma(x)}, & x > 0, \end{cases} \quad (6)$$

where $\Gamma(x, \lambda) \triangleq \int_\lambda^\infty e^{-t} t^{x-1} dt, x > 0, \lambda \geq 0$, is incomplete $\Gamma$-function, and $\Gamma(x) = \Gamma(x, 0)$ is $\Gamma$-function. One can prove that when $x$ is an integer, $\tilde{F}_\lambda(x) \equiv F_\lambda(x)$. So, we can use

$\tilde{F}_\lambda(x)$ as a proxy of $F_\lambda(x)$ to solve the original optimization problem. Specifically, we solve the following problem

$$(\hat{\beta}, \hat{\alpha}, \hat{\gamma}) = \underset{(\beta,\alpha,\gamma)}{\operatorname{argmax}}\, \tilde{l}_n(\beta, \alpha, \gamma)$$
$$= \underset{\beta,\alpha,\gamma}{\operatorname{argmax}} \sum_{i=1}^n \sum_{j=1}^{m_i} \sum_{k=1}^K I(Y_{ij} = k) \log\left[\tilde{F}_{\mu_{ij}(\beta,\Lambda_0)}(\gamma_k) - \tilde{F}_{\mu_{ij}(\beta,\Lambda_0)}(\gamma_{k-1})\right]. \quad (7)$$

A profile likelihood approach is used for estimating variances $pl_n(\beta, \alpha) = \max_\gamma \tilde{l}_n(\beta, \alpha, \gamma)$.

An R package 'opr' was developed to implement the algorithm and statistical inference for the proposed model and inference methods. The package 'opr' is publicly available at https://github.com/ActionSafe/FASTcure.

## 3  Simulation Study

Three simulation studies were conducted to evaluate the performance of the proposed method under real-world scenarios. In the following simulations, we set knots number $m_n$ to 2 and degree $l$ to 3 for the I-splines. We choose $3n^{-1/2}$ for perturbation constant $h_n$ and $1 \times 10^{-6}$ for (absolute and relative) convergence threshold.

### 3.1  Simulation 1

In the first simulation study, we considered two different scenarios to assess the behavior of the proposed method with finite sample sizes: the mean function is $\Lambda_1(t) = 15\log(1 + 0.7t) \cdot \exp(\beta_{11} X_1 + \beta_{12} X_2)$ for Scenario 1 and $\Lambda_2(t) = 3t \cdot \exp(\beta_{21} X_1 + \beta_{22} X_2)$ for Scenario 2. Here, $\beta_{11} = -1, \beta_{12} = 1, \beta_{21} = 1, \beta_{22} = 0$. $X_1$ (continuous) and $X_2$ (discrete) are independent $N(0, 1)$ and Binom(0.5) random variables. The follow-up times were randomly generated (independent of the recurrent event process): the maximum follow-up time $\tau$ was set to 10, and the number of follow-ups $m_i$ was drawn with equal probability from the integers 1 to 6. Then, $m_i$ samples were drawn from Unif(0, $\tau$) (rounded to one decimal), and duplicate values were removed. This guarantees a minimum follow-up interval of 0.1. For Scenario 1, the cut points were set as $\gamma_1 = (1, 3, 8)^T$, with $K_1 = 4$ levels. For Scenario 2, the breakpoints were $\gamma_2 = (3, 10)^T$, with $K_2 = 3$ levels. Each scenario was repeated 1,000 times for sample sizes $n = 200, 400, 800$.

For most simulated data, the BFGS algorithm converged within 100 iterations. Table 1 presents the simulation results for two scenarios under the known breakpoints. In the table, "Bias" represents the average bias when estimating covariate coefficients or the baseline function, "SD" represents the empirical standard deviation of the estimator, "SE" represents the average of the estimated standard errors, and "CP" indicates the coverage proportion of 95% confidence intervals. Table 2 presents the simulation results when the cut points are unknown. From the results in Table 1, it can be seen that our method achieves minimal parameter estimation bias and accurate standard errors with a small sample size of 200. The biases decrease as n increases. The results in Table 2 demonstrate that our proposed pseudo-likelihood function provides accurate estimates in the case of unknown cut points. Figures 1 and 2 display estimation results for $\Lambda_1(t)$ and $\Lambda_2(t)$. It shows that the I-spline estimator performs well when the cut points are known. When there is no information about cut points, a larger sample is needed to provide an accurate estimation.

**Table 1.** Results of simulation 1 with known cut points.

| | | Scenario 1 | | | | Scenario 2 | | | | |
|---|---|---|---|---|---|---|---|---|---|---|
| n | True | Bias | SD | SE | CP% | True | Bias | SD | SE | CP% |
| 200 | $\beta_{11} = 1.0$ | 0.006 | 0.091 | 0.087 | 95.0 | $\beta_{21} = 1.0$ | 0.007 | 0.081 | 0.080 | 94.8 |
| | $\beta_{12} = -1.0$ | −0.006 | 0.051 | 0.051 | 95.3 | $\beta_{22} = 0.0$ | <0.001 | 0.045 | 0.045 | 94.5 |
| | $\Lambda_1(2.5) = 15.2$ | 0.089 | 0.994 | 0.975 | 95.1 | $\Lambda_2(2.5) = 7.5$ | −0.013 | 0.44 | 0.448 | 95.4 |
| | $\Lambda_1(5.0) = 22.6$ | 0.098 | 1.330 | 1.315 | 95.8 | $\Lambda_2(5.0) = 15.0$ | −0.009 | 0.782 | 0.802 | 94.2 |
| | $\Lambda_1(7.5) = 27.5$ | 0.095 | 1.559 | 1.535 | 95.6 | $\Lambda_2(7.5) = 22.5$ | −0.015 | 1.130 | 1.153 | 95.4 |
| 400 | $\beta_{11} = 1.0$ | 0.003 | 0.061 | 0.061 | 94.9 | $\beta_{21} = 1.0$ | 0.002 | 0.057 | 0.056 | 94.8 |
| | $\beta_{12} = -1.0$ | −0.003 | 0.036 | 0.036 | 94.9 | $\beta_{22} = 0.0$ | −0.001 | 0.031 | 0.032 | 94.8 |
| | $\Lambda_1(2.5) = 15.2$ | 0.036 | 0.677 | 0.678 | 95.5 | $\Lambda_2(2.5) = 7.5$ | 0.013 | 0.316 | 0.315 | 95.7 |
| | $\Lambda_1(5.0) = 22.6$ | 0.034 | 0.925 | 0.914 | 95.3 | $\Lambda_2(5.0) = 15.0$ | 0.022 | 0.569 | 0.564 | 94.3 |
| | $\Lambda_1(7.5) = 27.5$ | 0.046 | 1.089 | 1.069 | 95.6 | $\Lambda_2(7.5) = 22.5$ | 0.049 | 0.826 | 0.81 | 94.6 |
| 800 | $\beta_{11} = 1.0$ | 0.002 | 0.043 | 0.043 | 95.4 | $\beta_{21} = 1.0$ | <0.001 | 0.038 | 0.040 | 95.7 |
| | $\beta_{12} = -1.0$ | −0.001 | 0.025 | 0.025 | 94.9 | $\beta_{22} = 0.0$ | <0.001 | 0.022 | 0.022 | 95.3 |
| | $\Lambda_1(2.5) = 15.2$ | 0.035 | 0.491 | 0.478 | 94.9 | $\Lambda_2(2.5) = 7.5$ | 0.004 | 0.229 | 0.222 | 94.4 |
| | $\Lambda_1(5.0) = 22.6$ | 0.036 | 0.658 | 0.645 | 94.8 | $\Lambda_2(5.0) = 15.0$ | 0.009 | 0.407 | 0.397 | 94.7 |
| | $\Lambda_1(7.5) = 27.5$ | 0.039 | 0.769 | 0.755 | 94.2 | $\Lambda_2(7.5) = 22.5$ | 0.011 | 0.578 | 0.571 | 95.9 |

**Table 2.** Results of simulation 1 with unknown cut points.

| | | Scenario 1 | | | | Scenario 2 | | | | |
|---|---|---|---|---|---|---|---|---|---|---|
| $n$ | True | Bias | SD | SE | CP% | True | Bias | SD | SE | CP% |
| 200 | $\beta_{11} = 1.0$ | 0.001 | 0.087 | 0.086 | 95.6 | $\beta_{21} = 1.0$ | −0.003 | 0.077 | 0.079 | 95.2 |
| | $\beta_{12} = -1.0$ | <0.001 | 0.050 | 0.052 | 95.4 | $\beta_{22} = 0.0$ | <0.001 | 0.046 | 0.044 | 94.4 |
| | $\Lambda_1(2.5) = 15.2$ | 0.972 | 1.570 | 1.800 | 96.2 | $\Lambda_2(2.5) = 7.5$ | 0.406 | 1.056 | 1.042 | 95.8 |
| | $\Lambda_1(5.0) = 22.6$ | 1.452 | 2.246 | 2.645 | 96.8 | $\Lambda_2(5.0) = 15.0$ | 0.833 | 2.083 | 2.032 | 95.0 |
| | $\Lambda_1(7.5) = 27.5$ | 1.791 | 2.711 | 3.236 | 96.6 | $\Lambda_2(7.5) = 22.5$ | 1.281 | 3.094 | 3.034 | 95.8 |
| 400 | $\beta_{11} = 1.0$ | −0.001 | 0.060 | 0.061 | 95.1 | $\beta_{21} = 1.0$ | −0.001 | 0.055 | 0.056 | 94.8 |
| | $\beta_{12} = -1.0$ | 0.002 | 0.036 | 0.037 | 95.2 | $\beta_{22} = 0.0$ | <0.001 | 0.029 | 0.031 | 96.6 |
| | $\Lambda_1(2.5) = 15.2$ | 0.639 | 1.054 | 1.240 | 96.0 | $\Lambda_2(2.5) = 7.5$ | 0.257 | 0.731 | 0.718 | 95.0 |
| | $\Lambda_1(5.0) = 22.6$ | 0.960 | 1.510 | 1.822 | 96.2 | $\Lambda_2(5.0) = 15.0$ | 0.498 | 1.439 | 1.397 | 94.4 |
| | $\Lambda_1(7.5) = 27.5$ | 1.190 | 1.824 | 2.228 | 96.4 | $\Lambda_2(7.5) = 22.5$ | 0.767 | 2.147 | 2.083 | 93.8 |
| 800 | $\beta_{11} = 1.0$ | −0.004 | 0.043 | 0.043 | 94.0 | $\beta_{21} = 1.0$ | −0.001 | 0.039 | 0.039 | 95.2 |
| | $\beta_{12} = -1.0$ | 0.002 | 0.027 | 0.026 | 94.2 | $\beta_{22} = 0.0$ | −0.001 | 0.021 | 0.022 | 96.8 |
| | $\Lambda_1(2.5) = 15.2$ | 0.348 | 0.723 | 0.858 | 96.8 | $\Lambda_2(2.5) = 7.5$ | 0.098 | 0.483 | 0.495 | 96.2 |
| | $\Lambda_1(5.0) = 22.6$ | 0.544 | 1.041 | 1.262 | 96.8 | $\Lambda_2(5.0) = 15.0$ | 0.189 | 0.933 | 0.963 | 97.2 |
| | $\Lambda_1(7.5) = 27.5$ | 0.666 | 1.248 | 1.542 | 96.8 | $\Lambda_2(7.5) = 22.5$ | 0.271 | 1.387 | 1.434 | 96.6 |

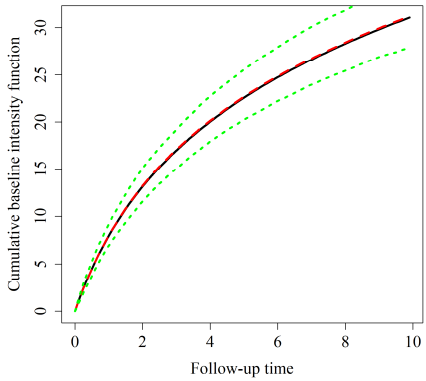

(**a**) Scenario 1, n=200

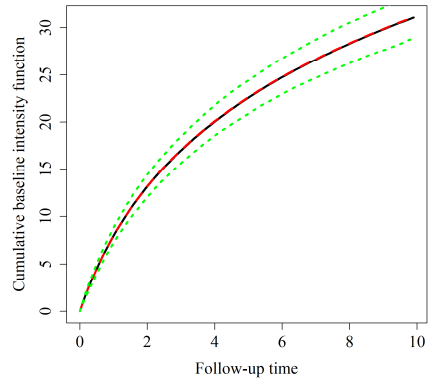

(**b**) Scenario 1, n=400

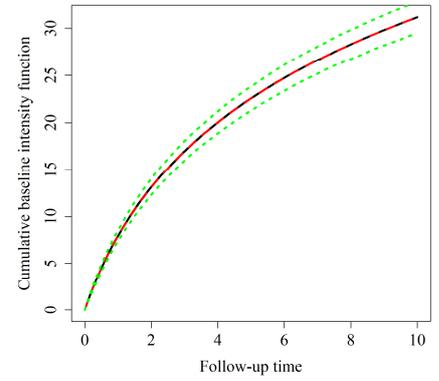

(**c**) Scenario 1, n=800

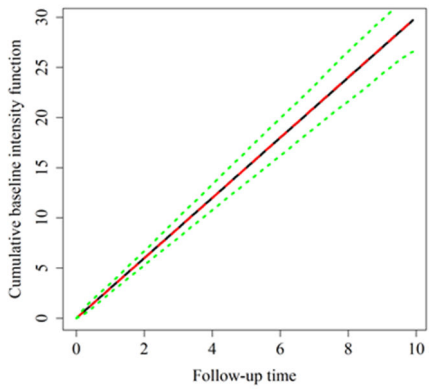
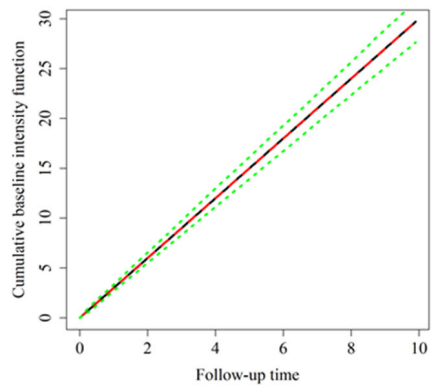
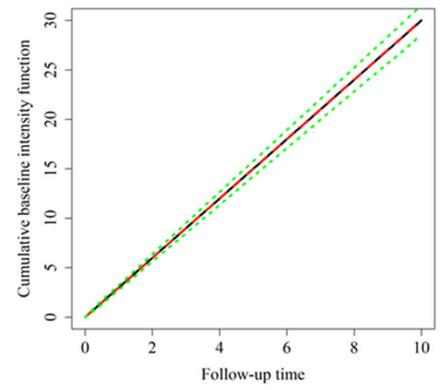

(**d**) Scenario 2, n=200      (**e**) Scenario 2, n=400      (**f**) Scenario 2, n=800

**Figure 1.** Results of simulation 1 on estimating the baseline mean functions with known cut points. The solid and dashed curves show the true values and averaged estimates, respectively, where each average is based on 1000 replicates.

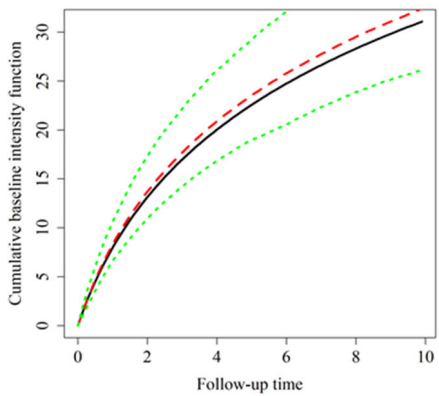
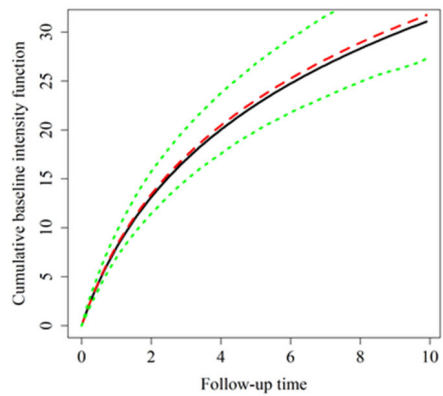
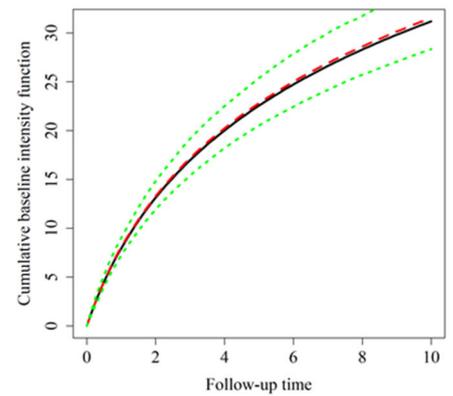

(**a**) Scenario 1, n=200      (**b**) Scenario 1, n=400      (**c**) Scenario 1, n=800

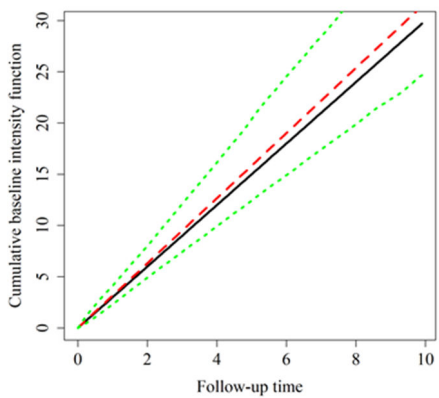
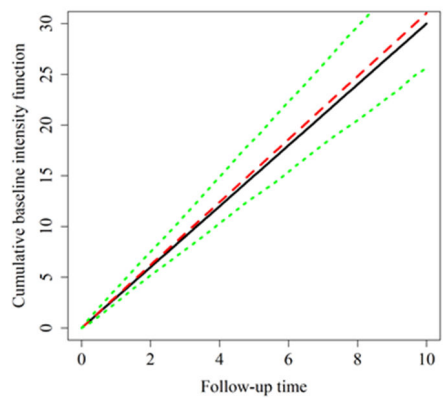
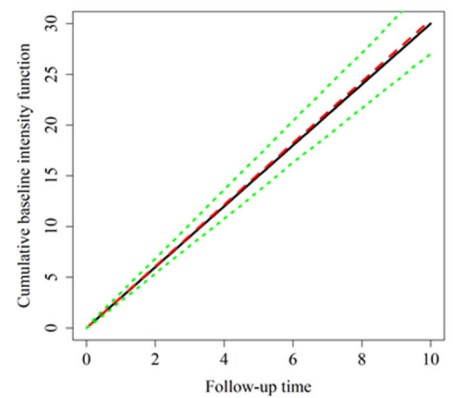

(**d**) Scenario 2, n=200      (**e**) Scenario 2, n=400      (**f**) Scenario 2, n=800

**Figure 2.** Results of simulation 1 for the estimation of the baseline mean functions with unknown cut points. The solid and dashed curves show the true values and averaged estimates, respectively, where each average is based on 1000 replicates.

## 3.2 Simulation 2

In the second simulation, two scenarios were considered to test proposed model's robustness to model assumptions. In the first scenario, we introduced individual frailty variable $Z$,

$$\lambda_i(t|X_i, Z) = Z\lambda_0(t)\exp(\beta^T X_i),$$

$Z$ follows a Gamma distribution with $E(Z)=1$. We tested two cases: Var($Z$) = 0.01 and Var($Z$) = 0.1. In the second scenario, we introduced a Box-Cox transformation to the mean function: $E(N_i(t)|X_i) = g\{\Lambda_0(t)\exp(\beta^T X_i)\}$, where $g(x) = [(1+x)^\rho - 1]/\rho$. We also tested the model's performance under two different choices of $\rho$: 1.05 and 1.10. The model's performance decreases as Var($Z$) and $\rho$ increase. The simulation results are summarized in Table 3.

**Table 3.** Results of simulation 2 with sample size n=400.

| Model | | | True | Cut points known | | | | Cut points unknown | | | |
|---|---|---|---|---|---|---|---|---|---|---|---|
| | | | | Bias | SD | SE | CP% | Bias | SD | SE | CP% |
| Frailty | $Z$~Gamma , Var($Z$)=0.01 | | $\beta_{11} = 1.0$ | 0.000 | 0.088 | 0.086 | 94.8 | −0.004 | 0.088 | 0.087 | 94.8 |
| | | | $\beta_{12} = -1.0$ | 0.001 | 0.055 | 0.051 | 93.8 | 0.007 | 0.056 | 0.052 | 93.4 |
| | $Z$~Gamma , Var($Z$)=0.1 | | $\beta_{11} = 1.0$ | −0.058 | 0.119 | 0.084 | 78.0 | −0.068 | 0.124 | 0.090 | 78.2 |
| | | | $\beta_{12} = -1.0$ | 0.050 | 0.075 | 0.049 | 71.4 | 0.068 | 0.074 | 0.054 | 68.0 |
| Transformation | $\rho = 1.05$ | | $\beta_{11} = 1.0$ | 0.056 | 0.084 | 0.087 | 91.4 | 0.045 | 0.084 | 0.086 | 92.4 |
| | | | $\beta_{12} = -1.0$ | −0.048 | 0.052 | 0.051 | 84.6 | −0.044 | 0.052 | 0.052 | 87.8 |
| | $\rho = 1.1$ | | $\beta_{11} = 1.0$ | 0.093 | 0.089 | 0.089 | 83.0 | 0.087 | 0.089 | 0.087 | 84.6 |
| | | | $\beta_{12} = -1.0$ | −0.092 | 0.052 | 0.052 | 60.6 | −0.088 | 0.055 | 0.053 | 61.6 |

**Table 4.** Results of simulation 3 with sample size n=400.

|  |  | Scenario 1 | | | | Scenario 2 | | | |
| --- | --- | --- | --- | --- | --- | --- | --- | --- | --- |
| Model | True | Bias | SD | SE | CP% | True | Bias | SD | SE | CP% |
| Proposed | $\beta_{11}=1.0$ | 0.006 | 0.091 | 0.087 | 95.0 | $\beta_{21}=1.0$ | 0.009 | 0.080 | 0.080 | 94.9 |
|  | $\beta_{12}=-1.0$ | −0.006 | 0.051 | 0.051 | 95.3 | $\beta_{22}=0.0$ | 0.001 | 0.046 | 0.045 | 93.9 |
| AEE | $\beta_{11}=1.0$ | −0.726 | 0.091 | 0.086 | 00.0 | $\beta_{21}=1.0$ | −0.709 | 0.078 | 0.083 | 00.0 |
|  | $\beta_{12}=-1.0$ | 0.723 | 0.051 | 0.049 | 00.0 | $\beta_{22}=0.0$ | −0.001 | 0.050 | 0.048 | 94.2 |
| CLMM | $\beta_{11}=1.0$ | 0.277 | 0.219 | 0.241 | 81.8 | $\beta_{21}=1.0$ | 0.452 | 0.223 | 0.238 | 55.0 |
|  | $\beta_{12}=-1.0$ | −0.287 | 0.135 | 0.144 | 50.6 | $\beta_{22}=0.0$ | −0.007 | 0.121 | 0.140 | 96.8 |

## 3.3 Simulation 3

We also conducted comparative studies with standard methods, including the AEE for panel count data and the CLMM for longitudinal ordinal data. All methods analyzed data from Simulation 1 (n=400, no cut points information available).

To generate data from AEE, we modify the baseline mean function to $\log(1+0.7t)$, now datasets are generated from Poisson process with mean function $\log(1+0.7t) \cdot \exp(\beta_{11} X_1 + \beta_{12} X_2)$. To generate data from CLMM, we use the following latent continuous model:

$$y_{it}^* = x_{it}^T \beta + v_i + \epsilon_{it},$$

where $v_i$ are independent and identically distributed $N(0, \sigma_v^2)$, $t = 1, \ldots, 5$ are follow-up times, $\epsilon_{it}$ are distributed as logistic with location 0 and scale 1. we observed ordinal responses $y_{it}$:

$$y_{it} = \begin{cases} 1, & y_{it}^* \leq \gamma_1 \\ 2, & \gamma_1 < y_{it}^* \leq \gamma_2 \\ \vdots & \\ K, & y_{it}^* > \gamma_{K-1} \end{cases},$$

we set K = 4 and $\gamma = (-1, 0, 1)^T$ for the simulation study.

As shown in Table 4, both AEE and CLMM yield biased estimation and under-covered confidence intervals. The poor performance of AEE is caused by naively treating the ordinal responses as true counts of event of interest. While CLMM doesn't account for the baseline mean function. Additional results on comparison of prediction accuracy and mean squared error are listed in Table 5 and Table 6, these results also support a better performance of the proposed methods.

Table 5. Comparison of the prediction accuracy (%)

|  | Scenario 1, n=400 | | | | Scenario 2, n=400 | | |
| --- | --- | --- | --- | --- | --- | --- | --- |
|  | Level 1 | Level 2 | Level 3 | Level 4 | Level 1 | Level 2 | Level 3 |
| Proposed | 51.7 | 52.0 | 66.8 | 90.4 | 71.9 | 77.2 | 90.2 |
| AEE | 41.9 | 29.3 | 24.3 | 19.4 | 44.1 | 37.8 | 27.6 |
| CLMM | 55.8 | 00.0 | 05.0 | 79.6 | 34.0 | 18.1 | 68.4 |

NOTE: Each simulation dataset was split into train (70%) and test (30%) datasets. All methods were trained on the train datasets and performs prediction on the test datasets.

Table 6. Comparison of mean squared errors (MSE)

|  | Data generation model | | | |
| --- | --- | --- | --- | --- |
| Methods | Proposed | AEE | CLMM | Average |
| Proposed | 0.392 | 1.706 | 2.134 | 1.411 |
| AEE | 3.215 | 0.602 | 1.431 | 1.749 |
| CLMM | 2.341 | 1.276 | 1.621 | 1.746 |

NOTE: We used three different models to generate data and each method was fitted on all datasets to get the MSE. Average of the MSEs for each method under three scenarios of data generation is listed in the last column.

# 4 Application to Medication Non-adherence Recurrence

We use proposed method to evaluate factors that could influence patients' medication adherence in the Sequenced Treatment Alternatives to Relieve Depression the Sequenced Treatment Alternatives to Relieve Depression (STAR*D) trial. This trial was a phase-IV multi-site, multi-stage randomized clinical trial to compare various treatment strategies for patients with non-psychotic major depressive disorder (Rush et al., 2004). The aim of the STAR*D study was to find the best subsequent treatment for subjects who failed to achieve adequate response to an initial antidepressant treatment (citalopram, Level 1 treatment). The STAR*D trial enrolled 4,041 outpatients with non-psychotic depression at 23 psychiatric and 18 primary care sites and obtained 80,820 observations in total. After excluding the unreasonable observation values and incomplete observations, the final data set for our analysis consisted of 1,967 patients with 7,214 observations and the maximum follow-up time up to 161 days.

Our interest is to assess the association between various baseline factors and the patient's medication adherence in Level 2 treatment. In this study, medication adherence is assessed by a categorical variable collected at approximately week 2, 4, 6, 9, 12 and 14 after entering Level 2 treatment. Patients were asked to recall 'How often missed medication since the last visit'. However, it is difficult for patients to recall the exact number of medication non-adherence, so only a ordinal variable is observed (called `L1-MAQ`). It takes 9 values: =1, never missing medicine; =2, rarely missing medicine; =3, sometimes; =4, half the time; =5, about half the time; =6, somewhat > half time; =7, very often; =8, nearly all the time; =9, all the time. During the follow, patients tend to not miss medicine (71.8%), there're few times when patients missed medicine for over half the time, so we combine L1-MAQ $\geq$ 4 into a single level of 4. After this combination, the new distribution of levels is: 1(71.8%), 2(21.5%), 3(4.4%) and 4(2.3%).

We use AIC and BIC criterion to determine optimal knots number and degree for I-spline. We select $m_n = 3$ and $l = 2$ under BIC criterion. We also compared our method with AEE and CLMM, the fitted covariate coefficients are shown in Table 7.

**Table 7.** Regression analysis of STARD trial

| Covarate | Proposed Model | | | AEE | | | CLMM | | |
|---|---|---|---|---|---|---|---|---|---|
| | Est | SE | P-value | Est | SE | P-value | Est | SE | P-value |
| FAMIN | 0.034 | 0.007 | <0.001 | 0.015 | 0.005 | 0.004 | 0.045 | 0.016 | 0.006 |
| SEX | -0.093 | 0.028 | <0.001 | -0.048 | 0.020 | 0.017 | -0.091 | 0.065 | 0.165 |
| White | -0.204 | 0.036 | <0.001 | -0.052 | 0.027 | 0.053 | -0.507 | 0.084 | < 0.001 |
| AGE | -0.012 | 0.001 | <0.001 | -0.004 | 0.001 | < 0.001 | -0.020 | 0.003 | < 0.001 |
| DEPPAR | 0.004 | 0.030 | 0.896 | 0.014 | 0.021 | 0.510 | -0.051 | 0.070 | 0.470 |
| QCBEG | -0.005 | 0.004 | 0.168 | -0.003 | 0.003 | 0.300 | -0.010 | 0.009 | 0.275 |

From Table 7, we conclude that family and friends' impact, sex, white race, and age are significant predictors for medication non-adherence, while parent history of depression and baseline QIDS are not significant. Majority of the results agree with the proposed method. Particularly, the results show that non-whites, males and younger patients are more likely to miss medication. AEE generally underestimates the impact of baseline covariates. Figure 3 shows estimated cumulative baseline intensity functions by two methods. Both methods find a significant increase of the cumulative intensity before day 120. While AEE shows a flattened tail of $\Lambda_0(t)$. The flattened tail of $\Lambda_0(t)$ and the underestimated coefficients may be due to naively treat ordinal response as true count thus causing AEE to underestimate the scale of event frequency.

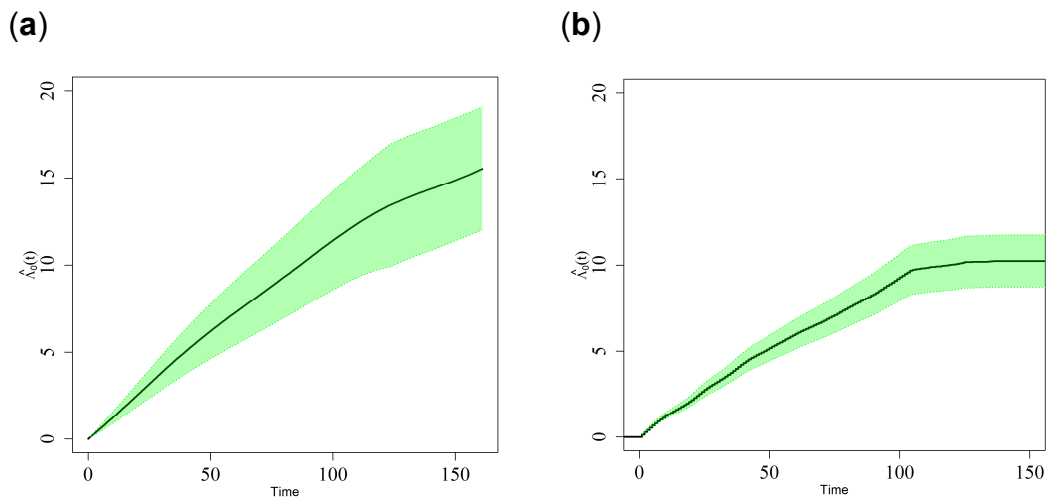

**Figure 3.** (**a**) cumulative baseline intensity function estimated by proposed method; (**b**) cumulative baseline intensity function estimated by AEE.

## 5 Discussion

In this study, we develop a model to analysis ordinal panel count data. This type of data occurs frequently in medical studies and brings difficulties to parameter estimation and inference because of its loss of information. Currently, no valid methods have been proposed to deal with this type of data and researchers usually view the data as either ordinary panel count data, or ordinal longitudinal data and use standard methods in these fields to conduct analysis. However, this can lead to severe bias and incorrect inference as shown in Simulation 3.

We handle these challenges by deriving an ordinal model based on proportional intensity assumption and introduce a continuous pseudo-likelihood estimator to solve vanishing gradients problem. Simulation studies have shown that both proposed sieve likelihood and pseudo-likelihood estimators work well in all Simulation 1 scenarios. In

Simulation 2, we show that performances can drop if we increase variances of random effects and shape parameter $\rho$ of Box-Cox transformation. That is a result of violation of Poisson process assumption and proportional dependent structure of covariates assumption. The Poisson assumption may be generalized by introducing random effects such as frailty in our model or deriving a similar augmented estimating equation. This would be a challenge, as conditional expectation in the E-step of EM algorithm requires a distribution assumption on $N(t)$ to compute when the exact number of events is not observed. It is also possible that the follow-up times are informative, our model can be adapted to this situation by a similar AEE approach. These topics are worthy of research in the future.

## Acknowledgements

This research was supported by the Noncommunicable Chronic Diseases-National Science and Technology Major Project (No. 2023ZD0506600), and the Capital's Funds for Health Improvement and Research (Grant numbers: 2024-1G-4251).

## Conflict of interest

The authors declare no conflict of interest.

## Data Availability Statement

The data will be made available by the authors on request.

## References

[1] Rush, A.J.; Fava, M.; Wisniewski, S.R.; Lavori, P.W.; Trivedi, M.H.; Sackeim, H.A.; Thase, M.E.; Nierenberg, A.A.; Quitkin, F.M.; Kashner, T.M.; et al. Sequenced Treatment Alternatives to Relieve Depression (STAR*D): Rationale and Design. *Control Clin Trials* **2004**, *25*, 119–142, doi:10.1016/s0197-2456(03)00112-0.

[2] Dziak, J.J.; Li, R.; Zimmerman, M.A.; Buu, A. Time-Varying Effect Models for Ordinal Responses with Applications in Substance Abuse Research. *Stat Med* **2014**, *33*, 5126–5137, doi:10.1002/sim.6303.

[3] Gebski, V.; Byth, K.; Asher, R.; Marschner, I. Recurrent Time-to-Event Models with Ordinal Outcomes. *Pharmaceutical Statistics* **2021**, *20*, 77–92, doi:10.1002/pst.2057.

[4] Barone, R.; Tancredi, A. Bayesian Inference for Discretely Observed Continuous Time Multi-State Models. *Statistics in Medicine* **2022**, *41*, 3789–3803, doi:10.1002/sim.9449.

[5] Sun, J.; Zhao, X. *Statistical Analysis of Panel Count Data*; Statistics for Biology and Health; Springer New York: New York, NY, **2013**; Vol. 80; ISBN 978-1-4614-8714-2.

[6] Dobson, A.J.; Barnett, A.G. *An Introduction to Generalized Linear Models*; Texts in statistical science; 3. ed.; Chapman & Hall, CRC: Boca Raton, Fla., **2008**; ISBN 978-1-58488-950-2.

[7] Guo, Y.; Sun, D.; Sun, J. Inference of a Time-Varying Coefficient Regression Model for Multivariate Panel Count Data. *Journal of Multivariate Analysis* **2022**, *192*, 105047, doi:10.1016/j.jmva.2022.105047.

[8] Xu, Y.; Zeng, D.; Lin, D.-Y. Proportional Rates Models for Multivariate Panel Count Data. *Biometrics* **2024**, *80*, ujad011, doi:10.1093/biomtc/ujad011.

[9] Zeng, D.; Lin, D.Y. Maximum Likelihood Estimation for Semiparametric Regression Models with Panel Count Data. *Biometrika* **2021**, *108*, 947–963, doi:10.1093/biomet/asaa091.

[10] Wang, X.; Ma, S.; Yan, J. Augmented Estimating Equations for Semiparametric Panel Count Regression with Informative Observation Times and Censoring Time. *Statistica Sinica* **2013**, *23*, 359–381.

[11] Li, Y.; He, X.; Wang, H.; Zhang, B.; Sun, J. Semiparametric Regression of Multivariate Panel Count Data with Informative Observation Times. *Journal of Multivariate Analysis* **2015**, *140*, 209–219, doi:10.1016/j.jmva.2015.05.014.

[12] Jiang, H.; Su, W.; Zhao, X. Robust Estimation for Panel Count Data with Informative Observation Times and Censoring Times. *Lifetime Data Anal* **2020**, *26*, 65–84, doi:10.1007/s10985-018-09457-7.

[13] Zhu, L.; Choi, S.; Li, Y.; Huang, X.; Sun, J.; Robison, L.L. Statistical Analysis of Clustered Mixed Recurrent-Event Data with Application to a Cancer Survivor Study. *Lifetime Data Anal* **2020**, *26*, 820–832, doi:10.1007/s10985-020-09500-6.

[14] Ge, L.; Hu, T.; Li, Y. Simultaneous Variable Selection and Estimation in Semiparametric Regression of Mixed Panel Count Data. *Biometrics* **2024**, *80*, ujad041, doi:10.1093/biomtc/ujad041.

[15] Ge, L.; Liang, B.; Hu, T.; Sun, J.; Zhao, S.; Li, Y. Variable Selection for Mixed Panel Count Data under the Proportional Mean Model. *Stat Methods Med Res* **2023**, *32*, 1728–1748, doi:10.1177/09622802231184637.

[16] Haubo, R., and Christensen, B. Cumulative Link Models for Ordinal Regression with the R Package Ordinal. **2018**, 1-46, https://api.semanticscholar.org/CorpusID:59572956.

[17] Rabe-Hesketh, S.; Skrondal, A.; Rabe-Hesketh, S. *Multilevel and Longitudinal Modeling Using Stata*; Multilevel and longitudinal modeling using Stata / Sophia Rabe-Hesketh, Anders Skrondal; Fourth edition.; Stata Press: College Station, **2022**; ISBN 978-1-59718-136-5.

[18] Liang, B.; Tong, X.; Zeng, D.; Wang, Y. Semiparametric Regression Analysis of Repeated Current Status Data. *Statistica Sinica* **2017**, *27*, 1079–1100.

[19] Hedeker, D.; Gibbons, R.D. A Random-Effects Ordinal Regression Model for Multilevel Analysis. *Biometrics* **1994**, *50*, 933–944, doi:10.2307/2533433.

[20] Ramsay, J.O. Monotone Regression Splines in Action. *Statistical Science* **1988**, *3*, 425–441.

[21] Fletcher, R. (Roger) *Practical Methods of Optimization*; Chichester; New York: Wiley, **1987**; ISBN 978-0-471-49463-8.